\documentclass[conference]{IEEEtran}

\usepackage{graphicx}
\usepackage{amsmath}
\usepackage{multirow}

\begin{document}

\title{ Aggregate AP Throughputs for Long File Transfers in a WLAN controlled by Inhomogeneous TCP Connections}

\author{Pradeepa BK and Joy Kuri \\ Centre for Electronics Design and Technology, \\
Indian Institute of Science, Bangalore. India. \\
{bpradeep,kuri}@cedt.iisc.ernet.in
}

\markboth{}
{ \MakeLowercase{}}

\maketitle

\begin{abstract}
The performance analysis of long file TCP controlled transfers in a WLAN in 
infrastructure mode is available in the present literature with one of the main 
assumptions being equal window size for all TCP connections.
In this paper, we extend the analysis to TCP-controlled long file uploads
and downloads
with different TCP windows. Our approach is based on simple Markov chain 
given in the paper \cite{astn_model:Kuriakose},
\cite{astn_model:Krusheel} with arbitrary window sizes.
We presented simulation results to show the accuracy of the analytical model.
\end{abstract}


\IEEEpeerreviewmaketitle

\section{Introduction}\label{sec:Introduction}
This paper is concerned with infrastructure mode WLANs that use the IEEE 802.11 
DCF mechanism. We are interested in analytical models 
for evaluating  the performance of TCP-controlled simultaneous uploads and
downloads where
each connection has arbitrary TCP windows size. 
A detailed analysis of the aggregate throughput of TCP download in a WLAN
for a single rate Access Point (AP) is given in 
\cite{astn_model:Kuriakose} by assuming constant TCP windows for all the links.
Similarly aggregate throughput of the AP is 
evaluated for the multi rate cases are in \cite{astn_model:Krusheel} and 
\cite{astn_model:pradeep_kuri}. However these works also consider either only  
download or upload with constant windows. Here, we 
consider both uploads and downloads along with different TCP windows.

We are motivated to study an analytical model for this scenario because of the clear 
understanding that it gives, and the useful insights that it can provide on the 
networks. 
The closed-form expression or numerical calculation procedures are useful because 
other features and capability can be built upon them. One such application, 
which we are studying now, is to utilize the results reported here in devising
a improved AP-STA association policy and improving the performance of the system.
Similarly if we can estimate the average upload and download throughput for the
AP as well as STAs.

Our approach is to model the number of STAs with ACKs and data packets in their
 MAC queues as  an embedded Discrete Time Markov Chain (DTMC), embedded at the 
instants of successful transmission events. We consider a successful 
transmission from the AP as a reward. This leads to viewing the aggregate TCP
throughput in the frame work of Renewal Reward theory given in \cite{astn_model:AKumar}  .

Our Contribution: We provide a simple approach to improve model of the 
aggregate throughput of the AP for long-lived TCP connection in terms of both
upload and download traffic with arbitrary maximum TCP receive window size in IEEE 
802.11 networks. We use the basic model and results presented in 
\cite{astn_model:Kuriakose}. We show 
that numerical results of our analytical model compare well with the 
simulation results. Simulations indicate that for the upload-download traffic
 scenario, our numerical evaluation of the analytical expression matches
 accurately with maximum error of $\pm$ 0.76 \% .

This paper is organized as follows: Section \ref{sec:Related_Work} outlines
related literature. In Section \ref{sec:System_Model} we 
state the system model and we discuss the modelling assumptions. In Section
\ref{sec:Analysis} we develop throughput analysis. In 
Section \ref{sec:Evaluation}, we present performance evaluation results. In 
Section \ref{sec:Discussion} we discuss the results.
Finally, the paper concludes in Section \ref{sec:Conclusion}.
\section{Related Work}\label{sec:Related_Work} 

The analytical work on this research area have been studied, considering 
saturated  and unidirectional traffic, i.e, either uplink or downlink in 
\cite{astn_model:bianchi}, \cite{astn_model:kumar} and \cite{astn_model:Cali} . All the above papers
assume that all the STAs are saturated, in other words they have all the time
packets to send to the AP. Our work deviates from this assumption. We consider 
the system with actual traffic load like TCP. 

All the related work that we are aware of assumes homogeneous TCP connections
in the sense that the maximum window size is the same for all connections.
\cite{astn_model:Kuriakose} and \cite{astn_model:Bruno4} propose a model for
 single rate AP-STA WLAN assuming the same maximum size of TCP window for all
 TCP connections. An extension of this work in \cite{astn_model:Krusheel}
 considers two rates of association with long file uploads from STAs to 
a local server and multirate case in \cite{astn_model:pradeep_kuri}.
\cite{astn_model:Bruno1} and \cite{astn_model:Bruno2} present another analysis
 of scenario of upload and download TCP-controlled file transfers with UDP traffic
  in a single cell infrastructure WLAN. They assumed equal TCP maximum window 
  size for all connections, that TCP receivers use undelayed ACKs, and showed 
  that the total TCP throughput is independent of the number of STAs in the 
  system. Also, upload and download transfers obtain an equal share of the 
  total throughput of the AP. 
The letter \cite{astn_model:Bruno3}, gives the average value analysis of TCP 
performance with upload and download traffic. First they provide expression 
for average number of active TCP stations. In \cite{astn_model:Onkar}, finite
buffer AP with TCP traffic in both upload and download direction is analysed 
with delayed and undelayed ACK cases. 

\cite{astn_model:Yu} provides an analysis for a given number of STAs and maximum
TCP receive window size by using well known $p$ persistent model proposed in 
\cite{astn_model:Cali}. However this \cite{astn_model:Yu} considers only download
traffic or upload traffic, not both together.

Another analysis HTTP traffic is in \cite{astn_model:Miorandi}, a queuing model
is proposed to
compute the mean session delay in the presence of short-lived TCP flows and studied 
the impact of TCP maximum congestion window size on this delay. The analysis
also extended to consider the delayed ACK technique.

\section{System Model}\label{sec:System_Model}

\begin{figure}
\centering
\includegraphics[scale=0.5]{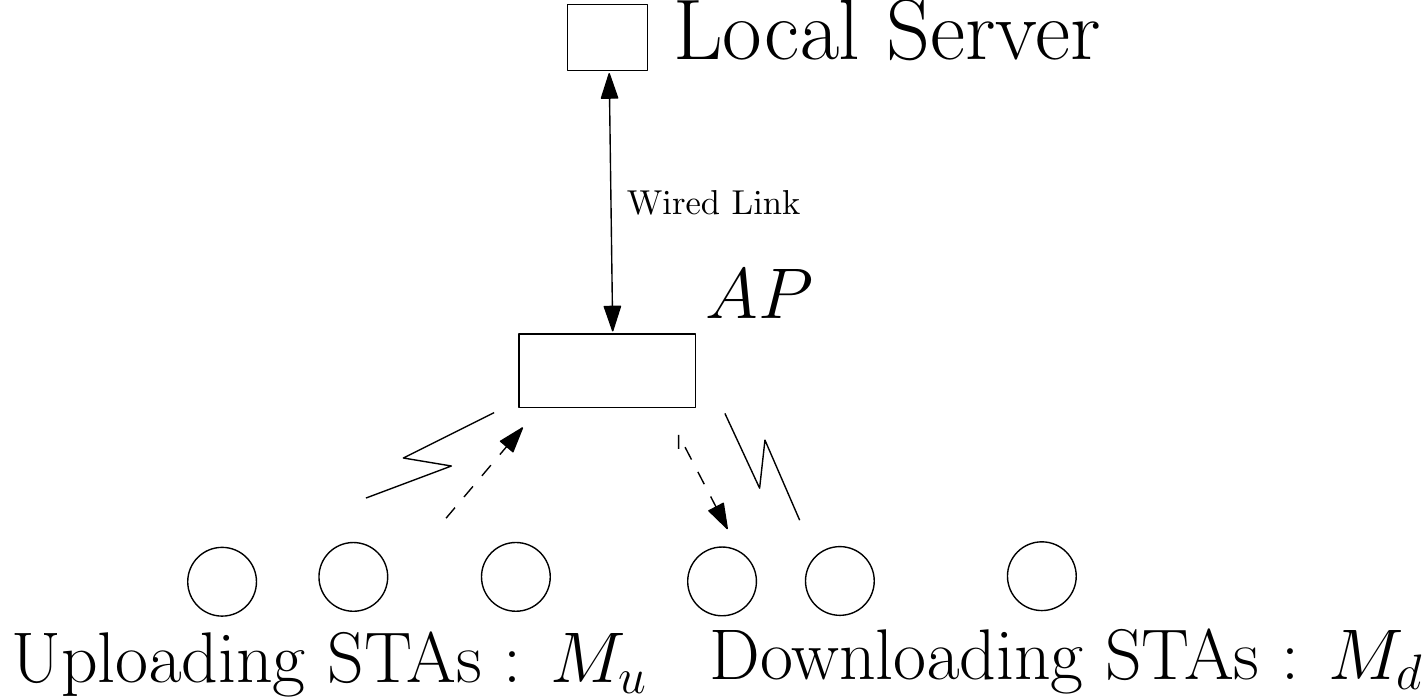} 
\caption{The network and traffic configurations. $M_u$ STAs are uploading 
files to a local server and $M_d$ STAs downloading files from the server 
through an AP.}
\label{fig:AP_STA}
\end{figure}

We consider a WLAN which has $M$ STAs associated with an AP as shown
in Figure \ref{fig:AP_STA}, and all the M STAs are associated at the same rate.
 We consider only TCP traffic. $M_u$ STAs are the 
senders in TCP uplink  connections and $M_d$ STAs are receivers in TCP downlink
connections. Thus, the AP sends either TCP ACK packets to the $M_u$ uploading STAs or
TCP data packets to $M_d$ downloading STAs. The arrows in Figure 
\ref{fig:AP_STA} show the direction of the data packets in the network. These are TCP 
links, there is also feedback traffic consisting of TCP-ACK packets
\footnotemark
\footnotetext[1]{We recall that every TCP receiver advertises a maximum window size.
In the absence of packet loss, a TCP sender's window will grow up to this advertised
window}.   We assume that $M$ is large. We will show later how big $M$ 
needs to be for our results to be applicable. 

Let $W_d$ the cumulative TCP advertised window size of all downloading STAs
and $ W_u $ the
cumulative TCP advertised window size seen by all uploading STAs. Let
$ W = W_u + W_d $ the sum of the maximum size of the TCP windows. All the nodes 
contend for the channel using the DCF mechanism as given in IEEE 802.11.
We assume that there are no link errors. Packets in the medium are lost only 
due to collisions. Because of the long file transfer scenario, we can assume
that TCP sources are operating
in Congestion Avoidance. Hence TCP startup transients can be ignored.  
Further, we assume that all the nodes use the RTS CTS mechanism while sending data 
packets and use basic access to send ACK packets. As soon as the
   station receives a data packet, it generates an ACK packet without any delay    
and it is enqueued at the MAC layer for transmission. We assume that all nodes 
 have sufficiently large buffers, so that packets are not lost due to buffer
 overflow. Also, TCP timeouts do not occur. The
 value of RTT is very small, since files are downloaded from a server 
 located on the LAN as shown in Figure \ref{fig:AP_STA}.
  
Thus, several TCP connections exist simultaneously and every 
station including the AP contends for the channel. Since no 
preference is given to the AP, and it has to serve all STAs, the AP becomes a 
bottleneck, and it is modelled as being backlogged permanently. The aggregate 
throughput of the AP is shared among all $ M $ stations.

\section{Analysis}\label{sec:Analysis}
The probability that the AP sends a TCP data packet to a downloading STA is
$p_d$ which
is equal to the ratio of the cumulative advertised window for downloading
connections (which we refer to as the ``cumulative download window'' henceforth),
to the sum of the cumulative
download and upload windows. Similarly, the probability the AP sends a TCP-ACK 
packet to an uploading STA is $p_u$ which is again, the ratio of the cumulative 
upload window, to the sum of the cumulative download and upload windows. i.e 
$ p_d = \frac{W_d}{W}$, and $ p_u = \frac{W_d}{W}$.

\begin{figure}
\centering
\includegraphics[scale=0.5]{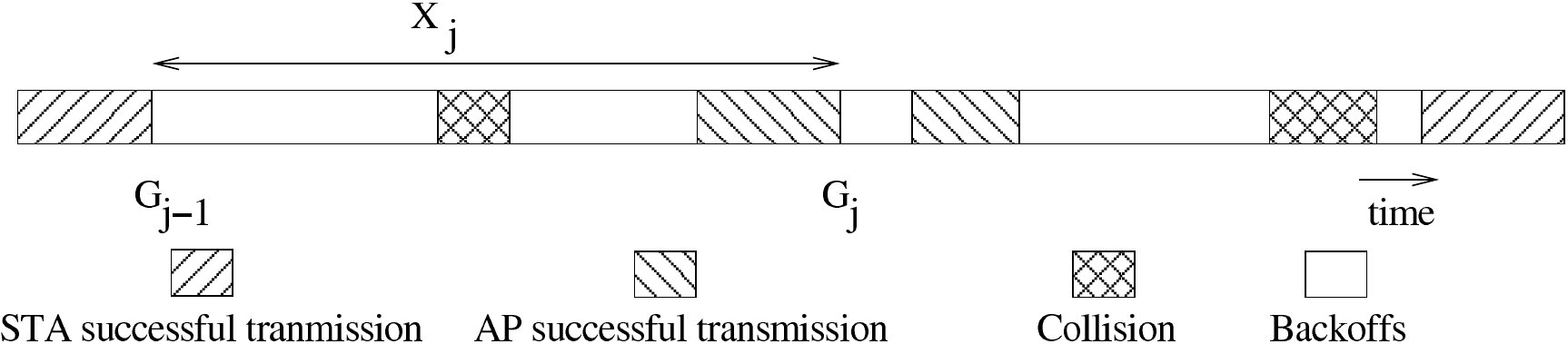} 
\caption{ Evolution of channel activity. $G_j$ are random time instants at which 
successful transmissions end. $X_k$ denotes the random duration of the $j^{th}$ 
contention cycle $[G_{j-1}, G_{j}) $. Each contention cycle consists of one or more 
back off periods and collisions but ends with successful transmission.}
\label{fig:channel_activity}
\end{figure}

Figure \ref{fig:channel_activity} shows one possible sample path of the events on 
the wireless channel in the WLAN. The random epochs $ G_j $ indicate the end 
of the $ j^{th} $ successful transmission
from either the AP or one of the stations. We observe that most STAs have empty 
MAC queues, because, in order for many STAs to have TCP-ACK packets or TCP data 
packets,
the AP must have had a long run of successes in the channel contention -- and this
 is unlikely because no special priority is given to the AP. So when the 
AP succeeds in transmitting, the packet is likely to be for a STA with an empty
MAC queue.

\begin{figure}[ht]
\centering
\includegraphics[scale=.5]{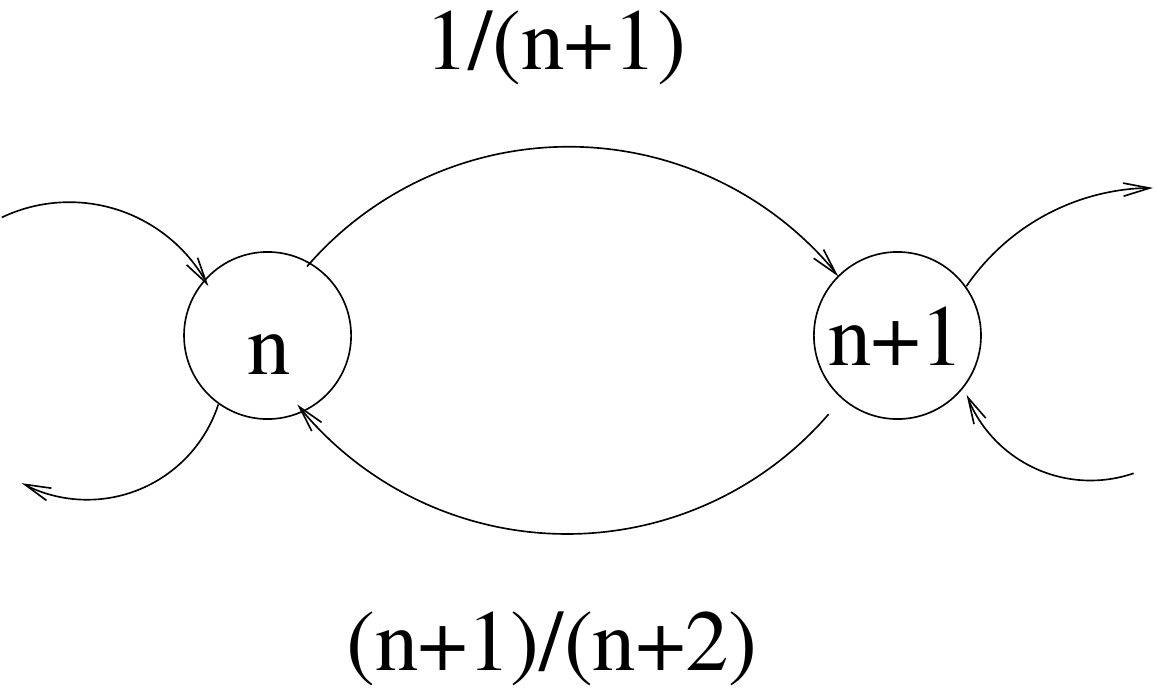}
\caption{Embedded Markov chain formed by the AP and $n$ stations 
associated to the AP at the same data rate. The state transition occurs
at the end of successful TCP data or TCP-ACK packet transmissions from the
contending nodes.}
\label{fig:MarkovChain}
\end{figure}

Let $ S_{j} $ be the number of STAs with nonempty MAC queues: if the 
STA is downloading, it has an ACK packet to send, or if it is uploading, then it 
has a TCP data packet to be sent. If there are $n$ nonempty STAs and a nonempty 
AP, each nonempty WLAN entity 
attempts to transmit with probability $ \beta_{(n+1)} $, where $ \beta_{n+1} $ 
is the 
channel access probability under saturation with $ (n+1) $ WLAN entities as in 
\cite{astn_model:kumar}. So $ S_{j}$ evolves as a 
DTMC over the epochs $ G_j $. This allows us to 
consider $(S_{j} , G_j) $ as a Markov Renewal Sequence,
and $S(t)$ as a semi-Markov process.
We have the DTMC which is shown in Figure \ref{fig:MarkovChain}; 
transition probabilities are indicated as well. By inspection, we can say that 
the DTMC is irreducible. The Detailed Balanced Equation holds for properly chosen 
set of equilibrium probabilities. The detailed balanced equation (DBE) is

\begin{equation}
 \pi_n \frac{1}{(n+1)}= \pi_{n+1} \frac{(n+1)}{(n+2)} 
\label{eq:DBE}
\end{equation}

Here, $ \pi_n $ is the stationary distribution of the DTMC. From the set 
of equations given in \eqref{eq:DBE} and 
$ \sum _{n=0} ^{ \infty }\pi_n  = 1 $, the stationary distribution is

\begin{equation}
\begin{split}
 \pi_n = (n+1)  \frac{1}{(n!)} * \frac{1}{(2e)}  
\end{split}
\label{eq:stationary_dist}
\end{equation}	
	
Let $ X $ be the sojourn time in a state $ ( S_{j} ) $. 
Conditioning on various events (idle slot, collision or successful 
transmission) that can happen in the next time slot, the
following expression for the mean cycle length can be written\\
\begin{equation}
\begin{split}
& E_{n}X = P_{idle}(\delta + E_{n}X) + ( P_{sAP}  T_{sAP} )\\
& + ( P _{c} ( T_{c} + E_{n}X )+ ( P_{sSTA}  T_{sSTA} ) 
\label{eq:enx11}
\end{split}
\end{equation}

In the above expression \eqref{eq:enx11}, $ P_{idle} $ is the probability 
of the slot being idle. $ P_{sAP} $ is the probability that the AP wins the 
contention and transmits the data packet or TCP-ACK packet. $ T_{sAP}$ is the
average time spent by the AP in a successful transmission; we have
\begin{equation*}
T_{sAP} = \frac{W_d}{W} T_{Data} + \frac{W_u}{W} T_{Ack}
\end{equation*}

$ T_{Data} $ is the time taken by the AP or STA to transmit a data packet and
$ T_{Ack} $ is the time taken by the AP or STA to transmit an ACK packet. 

 Detailed expressions and explanations are provided in the Appendix.
In the above expression when two or more transmission attempts occur, 
we have a collision. The duration of the collision ($T_c$) is given by the duration 
of the longest transmission time, i.e., the lowest rate of transmission determines
the duration of collision. The duration of the collision is decided by either 
the duration of RTS transmission or duration of TCP-ACK packet, depending 
on the physical rates.

Let $ T_{colliRTS} $ denote the duration of a collision given that the 
RTS is the longest packet involved in the collision. Then we have 
\begin{equation*}
T_{colliRTS} = T_p + T_{PHY} + \frac{L_{RTS} }{ r_c } + T_{EIFS}
\end{equation*}
where the notation is defined in Table \ref{tab:parameter}
Similarly, let $ T_{colliTCP-ACK} $ denote the duration of a collision 
given that the TCP-ACK is the longest packet involved in the collision.
We have
\begin{equation*}
\begin{split}
 T_{colliTCP-ACK} = & T_p + T_{PHY} + \\
& \frac{ L_{MAC} + L_{IPH}  + L_{TCP-ACK} }{ r_d } + T_{EIFS}
\end{split}
\end{equation*}

From Table \ref{tab:parameter} in Appendix (see also table of parameters 
in \cite{astn_model:Onkar}) in the above equation, it is clear that value
of data 
rate and control rate decides the duration of collisions as follows.
$ T_{colliTCP-ACK} < T_{colliRTS} $ if the data rate is 11 Mbps and 
$ T_{colliTCP-ACK} > T_{colliRTS} $ if the data rate is 5.5 Mbps or 2 Mbps.
The various possibilities of collisions can now be listed as follows.
 
In the first case, $ T_c $ equals $ T_{colliRTS} $ under two conditions:
(i) AP data packet transmission collides with either uploading 
or downloading STA (ii) AP TCP-ACK packet transmission collides with
uploading STA. In the second case, $T_c$ equals $ T_{colliTCP-ACK}$ when
AP TCP-ACK packet transmission collides with downloading STA Further 
explanation is given in Table \ref{table:collision} in Appendix.

 In the above expression, various probabilities have been
obtained by considering the events and using channel access probability $\beta _{N+ 1} $,
when there are $(N + 1)$ contending nodes.

From Equation \eqref{eq:enx11} we have 
\begin{equation}
E_{n}X = \frac{ P_{idle} + P_{sAP} T_{sAP} + P_{c}T_{c} + P_{sSTA} T_{sSTA} }
 {1- P_{idle} -P_{sAP} - P_{c} - P_{sSTA} }
\label{eq:enx}
\end{equation}

Calculations of probabilities and times in \eqref{eq:enx} are shown in the Appendix.
The mean reward for a cycle is obtained as follows. We are interested in finding the 
long run time average of successful transmissions from the AP. This leads 
to Markov regenerative analysis or the renewal reward theorem approach. 
To get the mean renewal cycle length, we can 
use the mean sojourn time given in Equation \eqref{eq:enx}. The mean reward in a cycle can 
be obtained as follows. A reward of 1 is earned when the AP transmits either a TCP data 
packet or an ACK packet successfully by winning the channel. The probability of the AP winning
the channel is  $ \frac{1}{(n + 1)}$. Hence the semi Markov 
process exits the state $ (n)$ with probability $ \frac{1}{(n + 1)}$. A reward of 0 is earned with the probability $ ( 1- \frac{1}{(n + 1)} )$. Therefore, the expected reward is  $ \frac{1}{(n + 1)}$. So this results in aggregate throughput of the AP with both upload and download.

Hence the aggregate TCP throughput  in this case can be 
calculated as\\
\begin{equation}	
 \Phi_{AP-TCP} = \frac{\Sigma^{\infty}_{n=0} \pi_n (\frac{1}{n+1}) 
}{\Sigma^{\infty}_{n=0} \pi_n  E_{n}X}  
\label{eq:ap_thpt}
\end{equation}

Further, we can consider only upload throughput or download throughput by 
changing the assignment of rewards. If we count a reward of 1 when the AP 
transmits only TCP data packet and reward of 0 else (even though the AP wins
 the channel and transmits TCP Ack packet), we can obtain 
the aggregate download throughput $ \Phi_{d} $. Similarly, if we count a 
reward of 1 when the AP transmits a TCP-ACK packet, we can get aggregate 
upload throughput $ \Phi_{u}$ .

\section{Evaluation}\label{sec:Evaluation}
To verify the accuracy of the model, we performed experiments using the Qualnet 4.5
network simulator \cite{astn_model:Qualnet}. We considered 802.11b physical data 
rates as 11, 5.5 and 2 Mbps. 
In Table \ref{table:mac_11b}, results are given for a few cases of this 
scenario, i.e. with different number of stations having different maximum size of
TCP receive windows. Table \ref{table:mac_11b} gives the number of downloading
and uploading STAs with TCP window sizes being 24, 20 and 14 packets, and the 
data rates being 11, 5.5 and 2 Mbps. Analytically calculated aggregate 
throughputs are listed against simulation results with 95\% confidence 
intervals for 30 runs in the right side columns of Table \ref{table:mac_11b}.     
\begin{table}[ht]
\centering 
\begin{tabular}{|c|c|c|c|c|c|c|c|c|} 
\hline 
Rate & \multicolumn{6}{|c|}{No.of STAs with windows size} &  \multicolumn{2}{|c|}{ Aggregate }\\ \cline{2-7} 
Mbps & \multicolumn{3}{|c|}{Downloading}  & \multicolumn{3}{|c|}{Uploading} &  \multicolumn{2}{|c|}{ Throughput [Mbps] }\\ \cline{2-9}
 & 24 & 20 & 16 & 24 & 20 &  16 & Analysis & Simulation \\  [0.2ex]
\hline 
\multirow{2}{*}{11} & 1	& 2	& 3	& 4	& 2	& 3 & 4.38  & 4.37$\pm$ 0.01 \\
				& 2	& 1  & 3	& 4	& 2  & 3 & 4.38  & 4.37 $\pm$ 0.01 \\ \hline
\multirow{2}{*}{5.5}& 3	& 2	& 1	& 4	& 2	& 3 & 3.04  & 3.04 $\pm$ 0.01\\
				& 4	& 3	& 2	& 1	& 3	& 2 & 3.04 & 3.04 $\pm$ 0.01 \\ \hline
\multirow{2}{*}{2}  & 3	& 2	& 4	& 3	& 1	& 2 & 1.5 & 1.5 $\pm$ 0.001 \\
				& 3	& 2	& 4	& 3	& 2	& 1 & 1.5 & 1.5 $\pm$ 0.001 \\ [1ex]
\hline 
\end{tabular}
\caption{Aggregate Throughput [\textit{Mbps}] of the AP by analysis and simulation
 for IEEE 802.11\textit{b} with TCP receive window sizes being 24,20 and 16 for 
 uploads and downloads} 
\label{table:mac_11b} 
\end{table}

In 802.11g, data rates are 54, 48, 36, 24, 18, 12 and 6 Mbits/s. Qualnet 4.5, is
configured to this mode by setting the channel frequency for 802.11a radio as 
2.4 GHz. In Table \ref{table:Mac_11g}, comparisons between analytical and 
simulation values are given for selected data rates to illustrate the accuracy 
of the analytical model. The values of maximum TCP receive windows are shown in 
the heading of the table as 24, 20 and 16 packets.
Number of STAs with these as TCP windows are listed for both upload and download links.
The rights side columns give aggregate throughput for the different values of data rates.
Simulation results presented with 95\%  confidence interval for 30 runs.


\begin{table}[ht]
\centering 
\begin{tabular}{|c|c|c|c|c|c|c|c|c|} 
\hline 
Rate & \multicolumn{6}{|c|}{No.of STAs with windows size} &  \multicolumn{2}{|c|}{ Aggregate }\\ \cline{2-7} 
Mbps & \multicolumn{3}{|c|}{Downloading}  & \multicolumn{3}{|c|}{Uploading} &  \multicolumn{2}{|c|}{ Throughput [Mbps] }\\ \cline{2-9}
 & 24 & 20 & 16 & 24 & 20 &  16 & Analysis & Simulation \\  [0.2ex]
\hline 
\multirow{2}{*}{54} & 1	& 2	& 3	& 4	& 2	& 3 & 22.61  & 22.5 $\pm$ 0.01 \\
				& 4	& 1  & 2	& 2	& 1  & 3 & 22.61  & 22.56 $\pm$ 0.01 \\ \hline
\multirow{2}{*}{48}	& 3	& 2	& 1	& 4	& 2	& 3 & 19.68  & 19.54 $\pm$ 0.01\\
				& 4	& 3	& 2	& 1	& 3	& 4 & 19.68 & 19.53 $\pm$ 0.01 \\ \hline
\multirow{2}{*}{36}	& 3	& 2	& 1	& 4	& 2	& 3 & 14.94  & 14.92 $\pm$ 0.01\\
				& 4	& 3	& 2	& 1	& 3	& 2 & 14.94 & 14.92 $\pm$ 0.01 \\ \hline
\multirow{2}{*}{12}	& 3	& 2	& 4	& 3	& 1	& 2 & 5.16 & 5.15 $\pm$ 0.001 \\
				& 3	& 2	& 1	& 3	& 2	& 4 & 5.16 & 5.14 $\pm$ 0.001 \\ [1ex]
\hline 
\end{tabular}
\caption{Comparison of Throughput [\textit{Mbps}] of the AP by analysis and simulation for IEEE 802.11\textit{g} with window sizes being 24, 20 and 16 for uploads and downloads}
\label{table:Mac_11g} 
\end{table}
As discussed in Section \ref{sec:Analysis}, the aggregate throughput of the AP 
is divided between aggregate upload and download in proportion to 
their maximum
TCP receive window sizes. The numerical evaluation of Equation 
\eqref{eq:ap_thpt} is given in Table \ref{table:down_up_thpt} for all, i.e., download
upload and aggregate throughput along with simulation results
with 95\%
confidence interval for 30 runs.
\begin{table}[ht]
\centering 
\begin{tabular}{|c|c|c|c|c|} 
\hline 

Rate & \multicolumn{2}{|c|}{AP Download}  & \multicolumn{2}{|c|}{AP Upload} \\ 
Mbps	& \multicolumn{2}{|c|}{Throughput [Mbps]}  & \multicolumn{2}{|c|}{Throughput [Mbps]} \\ \cline{2-5}
     & Analysis & Simulation & Analysis & Simulation \\  [0.2ex]
\hline 
\multirow{2}{*}{54} & 8.56  &	8.51	 $\pm$ 0.01	&	13.987	&	14.055 $\pm$ 0.01 \\	
				& 12.68 &	12.65 $\pm$ 0.01	&	9.935	&	9.9127 $\pm$ 0.01 \\ \hline
\multirow{2}{*}{48} & 8.074 &	8.016 $\pm$ 0.01	&	11.524	&	11.606 $\pm$ 0.01 \\
				& 11.011&	10.94 $\pm$ 0.01	&	8.603	&	8.6686 $\pm$ 0.01 \\ \hline
\multirow{2}{*}{36} & 6.13  &	6.12  $\pm$ 0.01	&	8.799	&	8.812 $\pm$ 0.01 \\
				& 9.24  &	9.23  $\pm$ 0.01	&	5.693	&	5.701 $\pm$ 0.01 \\ \hline
\multirow{2}{*}{12} & 3.027 &	3.021 $\pm$ 0.001	&	2.127	&	2.133 $\pm$ 0.001 \\
				& 2.173 &	2.164 $\pm$ 0.001	&	2.976	&	2.987 $\pm$ 0.001 \\ [1ex]
\hline 
\end{tabular}
\caption{Aggregate upload and download throughput [\textit{Mbps}] of the
 AP by analysis and simulation for IEEE 802.11\textit{g} with scenario presented in
  Table \ref{table:Mac_11g} } 
\label{table:down_up_thpt}
\end{table}

\section{Discussion}\label{sec:Discussion}
In this work, we presented an analytical model to obtain the aggregate 
throughput when several TCP-controlled long file upload and downloads are
 going on with arbitrary window sizes. Calculation of probabilities and 
 durations of all the events need to consider physical data rates and
  control rates. The analysis of throughput made use of only the fact that 
the TCP source operates in Congestion Avoidance.

We can notice that in Tables \ref{table:mac_11b} and \ref{table:Mac_11g}, 
for a specified data rate and control rate of transmission,
there is no discrepancy in the aggregate throughput (both upload and download)
with the number of STAs. It motivates us to use this model as a processor 
sharing model for arbitrarily
   arriving short file transfers. Every downloading STA will receive the service 
   rate of $ \Phi_{d} N_{d} $    (for uploading STA, $ \frac{ \Phi_u }{ N_u  }$ ) 
   as mentioned in \cite{astn_model:Onkar}.    In our simulation and numerical
    evaluation we used the 802.11b and 802.11g standards. However, our mathematical
expressions are independent of these standards; hence the model can be applied to 
any other standard that has different number of physical data rates. We assumed no 
packet losses and no channel errors; we need to address these by 
introducing link error probability. Also we assumed that RTT is negligible, 
which needs to be generalized. 

\section{Conclusion}\label{sec:Conclusion}
This paper developed a simple general analytical framework to obtain accurate 
closed-form expressions for the performance of the AP with long-lived TCP 
connections in IEEE 802.11 networks.
In this work, we have presented a model for the aggregate throughput of the 
AP by considering simultaneous TCP upload and download traffic for long files
with arbitrary 
TCP window sizes. We verified the correctness of the analytical model with the 
simulation results. These results show the accuracy of the model, with the 
maximum 
error being $\pm$ 0.76 \% . We considered single data rate of association. This approach 
can be extended to multiple rates but this makes the model more complicated and 
consequent state space expansion makes the calculation of stationary
probability distribution tedious. This is a particular limitation of our
approach. However the model can be used in addressing other performance
evaluation  questions of WLANs.


\appendix[ Expressions for probabilities and times discussed in section \ref{sec:Analysis} ]

%

\appendices

\begin{tabular}{ p{.7cm} p{7cm} }
$ P_{idle} $ & is the probability of the slot being idle. \\
 			  & $ =  (1 -\beta_{N+1})^{N+1} $ \\
 $ P _{sAP} $ & \text{is probability of AP wins the contention and}\\
 			  & \text{transmits the packet }.\\
 			  & $ =  \beta_{N+1} (1 - \beta_{N+1})^{N}$\\
 $ P _{sSTA} $ & \text{ is the probability of STA the contention  } \\
 			  & $ = n_i \beta _{N+1} (1- \beta _{N+1} )^N $	\\		  
 $ P _{c} $    & \text{ is the probability of the collision event  } \\
			  & $ = \beta _{N+1} ( 1 - (1- \beta _{N+1})^N )$ \\
$ T _{sAP} $  & is the mean time taken by the AP to send packet (TCP data or TCP-ACK) to an STA\\
			 & $ = \frac{ W_{d}}{ W } T _{Data} + \frac{ W_{u}}{ W }  T_{Ack}$ \\
$ T _{sSTA} $    &  is the duration of transmitting one TCP data packet or one TCP-ACK  \\
			  &   packet from STA including overhead  \\
	  		 & $ = \frac{ W_{u}}{ W } T _{Data} + \frac{ W_{d}}{ W }  T_{Ack}$ \\
$ T _{c} $ &  is the collision duration when the AP and    \\
			  & STAs are involved. \\
			  & $ =T_p + T_{PHY}+\frac{ L_{MAC}+L_{IPH}+L_{TCP-ACK} }{ r_d } + T_{EIFS}$ \\  
$ T_{Data} $ &  is the time taken by any node to transmit an TCP data packet. \\
			& $ = T_p + T_{PHY} + \frac{ L_{RTS}}{ r_c } + T_{SIFS} + T_{p} $ \\ 
			  & $ + T_{PHY} + \frac{ L_{CTS} }{ r_c }  + T_{SIFS} + T_p + T_{PHY} $\\
			  & $ + \frac{ L_{MAC} + L_{IPH} + L_{TCPH} + L_{TCP} }{ r_d } $ \\
			  & $ + T_{SIFS} + T_{p} + T_{PHY}  + \frac{ L_{ACK} } { r_c } + T_{DIFS} $ \\
$ T_{Ack} $ & is the time taken by any node to transmit an TCP ACK.\\
		& $ = T_p + T_{PHY} + L_{MAC} + \frac{ L_{IPH} + L_{TCP-ACK} }{r_i} $ \\
		  & $ + T_{SIFS} + T_{p} + T_{PHY} + \frac{ L_{ACK} }{ r_d } + T_{DIFS} $ \\ 
\end{tabular}

\begin{table}[ht]
\centering 
\begin{tabular}{|c|c|c|} 
\hline 
\hline
 Collision & AP(STA) sends Data & AP(STA) sends TCP ACK \\  [0.5ex]
\hline \hline 
 \multicolumn{3}{|c|}{Data rate is 11 Mbps} \\ \hline
Uploading STA & Length of RTS & Length of RTS \\ 
Downloading STA & Length of RTS  &  Length of TCP-ACK  \\ \hline
 \multicolumn{3}{|c|}{Data rate is either 2 or 5.5 Mbps} \\ \hline
Uploading STA &  Length of RTS & Length of TCP-ACK  \\
Downloading STA &  Length of TCP-ACK & Length of TCP-ACK  \\ [1ex] 
\hline 
\end{tabular}
\caption{ Duration of Collision between AP and STA or two STAs in channel with rate of 11, 5.5 and 2 Mbps and Control rate of 2 Mbps }
\label{table:collision} 
\end{table}
The values of these parameters are given in Table \ref{tab:parameter}

\begin{table}
\begin{tabular}{|c|c|c|c|}
\hline
 Parameters     			& Symbol		& 802.11b 	& 802.11g 	\\ \hline \hline
 Max PHY data rate			& $r_d$ 		& 11 Mbps		& 54 Mbps		\\ \hline
 Control rate				& $r_c$ 		& 2 Mbps		& 6 Mbps 		\\ \hline
 PLCP preamble time			& $T_p$ 		& $144\mu s$	& 			\\ \hline
 PHY Header time			& $T_{PHY}$ 	& $48  \mu s$	& $20 \mu s$	\\ \hline
 MAC Header size			& $L_{MAC}$	& 34 bytes	& 34 bytes	\\ \hline
 RTS Header size			& $L_{RTS}$	& 20 bytes	& 20 bytes	\\ \hline
 CTS Header size			& $L_{CTS}$	& 14 bytes	& 14 bytes	\\ \hline
 MAC ACK Header size		& $L_{ACK}$	& 14 bytes	& 14 bytes	\\ \hline
 IP Header size			& $L_{IPH}$	& 20 bytes	& 20 bytes	\\ \hline
 TCP Header size			& $L_{TCPH}$	& 20 bytes	& 20 bytes	\\ \hline
 TCP ACK Packet size 		& $L_{TCP-ACK}$& 20 bytes	& 20 bytes	\\ \hline
 TCP data payload size 		& $L_{TCP}$	& 1460 bytes	& 1460 bytes	\\ \hline
 System slot time			& $\delta$	& $20 \mu s$	& $9  \mu s$ 	\\ \hline
 DIFS time 				&$T_{DIFS}$	& $50 \mu s$	& $28 \mu s$  	\\ \hline
 SIFS time 				&$T_{SIFS}$	& $10 \mu s$	& $10 \mu s$  	\\ \hline
 EIFS time 				&$T_{EIFS}$	& $364 \mu s$	& $364 \mu s$  \\ \hline
 CWmin					& CWmin 		& 31 		& 15			\\ \hline
 CWmax					& CWmin 		& 1023 		& 1023		\\ \hline
\end{tabular}
\caption{Values of Parameters used in Analysis and Simulation} \label{tab:parameter}
\end{table} 




%




\end{document}